\documentclass[11pt]{article}
\def\d{\partial}
\def\db{\bar\partial}
\def\t{\theta}
\def\tb{\bar\theta}
\usepackage{oldgerm}
\input{amssym}
\def\be{\begin{equation}}
\def\ee{\end{equation}}

\newsavebox{\PSLASH}
\sbox{\PSLASH}{$p$\hspace{-1.8mm}/}

\begin{document}
\title{Higher Order and Boundary Scaling Fields in the Abelian Sandpile Model}
\author{S. Moghimi-Araghi\footnote{e-mail: samanimi@sharif.edu}, A. Nejati \footnote{e-mail: nejati$\_$a@sharif.edu}  \\
Department of Physics, Sharif University of Technology,\\ Tehran,
P.O.Box: 11365-9161, Iran} \date{} \maketitle

\begin{abstract}
The Abelian Sandpile Model (ASM) is a paradigm of self-organized
criticality (SOC) which is related to $c=-2$ conformal field theory.
The conformal fields corresponding to some height clusters have been
suggested before. Here we derive the first corrections to such
fields, in a field theoretical approach, when the lattice parameter
is non-vanishing and consider them in the presence of a boundary.
  \vspace{5mm}%
\end{abstract}

\section{Introduction}
Self-organized criticality is believed to be the underlying reason
for the scaling laws seen in a number of natural phenomena
\cite{Jensen}. Dynamics of a self-organized critical system is such
that the system naturally approaches its critical state and exhibits
long range orders and scaling laws. This means that, unlike most of
statistical models, without any fine-tuning of some parameters such
as temperature, the system reaches its critical point.

The concept was first introduced by Bak, Tang and Wiesenfeld
\cite{BTW}. In their paper they proposed the Abelian Sandpile Model
(ASM) as a model of self-organized criticality. Since then many
different models exhibiting the same phenomenon have been developed,
but still ASM is the simplest, most studied model, in which many
analytical results has been derived. For a good review see ref.
\cite{DharRev}.

Many exact results are derived in this theory. The first analytical
calculation, which paved the road for other analytical results, was
done by Dhar \cite{Dhar}. Probabilities of some specific clusters,
known as Weakly Allowed Clusters (WAC's), was then computed
\cite{MajDharJPhysA}. The simplest of these clusters is one-site
height one cluster. The probabilities of other one-site clusters
with height above 1 was computed in ref. \cite{Prizh}. Among other
analytical results one can mention the results on boundary
correlations of height variables and effect of boundary conditions
\cite{Ivashkevich94,Jeng05-1,Jeng05-2,Jeng04,Ruelle02}, on presence
of dissipation in the model \cite{Jeng05-2,Jeng04,PirRu04}, on field
theoretical approaches \cite{MaRu,Jeng05-2,MRR,JPR}, on finite size
corrections \cite{Ruelle02,MajDharJPhysA} and many other
results\cite{DharRev}.

On the other hand, the model has been related to some other lattice
models such as spanning trees, which correspond to the well-known
$c=-2$ conformal model. Mahieu and Ruelle \cite{MaRu} related ASM to
logarithmic conformal field theory through introducing scaling
fields corresponding to the Weakly Allowed Clusters (WACs), such as
one site height-1 cluster. This was done through comparison of
correlation functions in the two models, and the fields
corresponding to a list of simplest WAC's was obtained. Later, Jeng
\cite{Jeng05-1} introduced an elegant method to calculate such
fields for arbitrary WAC and showed that for all of these clusters
the the scaling dimension is two. Meanwhile, a more direct way to
show this correspondence was developed in ref. \cite{MRR}. The
benefit of this new method is that it comes from an action and is
not found merely by comparing correlation functions. This allows
some further investigations that was not possible before. The fields
associated with other one-site clusters are also derived
\cite{PirRu04,JPR,Jeng05-2}.

However, most of these results are obtained in the thermodynamic
limit; i.e., it is supposed that the size of the system, $L$, is
very large and the lattice spacing, $a$, is ignorable. Though some
results are obtained in situations where these conditions are not
satisfied, we would like to use the method introduced in ref.
\cite{MRR} to obtain scaling fields when we are far from
thermodynamic limit. To loose the first assumption ($L\rightarrow
\infty$) one should consider the finite size effects and to see the
effects of loosing the second condition, one may consider how the
scaling fields depend on the lattice spacing. Because of discrete
nature of the model, it is important to investigate this problem.

In this paper we first review the method introduced in \cite{MRR}
to derive scaling fields and then using the method, we derive the
higher order corrections to the first order of lattice spacing. In
the end we discuss the effect of boundary on the field derived
using the method of \cite{MRR}.

\section{Scaling Fields}
As mentioned before, a correspondence between ASM and other
well-known statistical models has been established, the most
significant of which are the connection to $q\rightarrow 0$ limit of
the q-state Potts model \cite{denspol}, spanning trees
\cite{MajDharJPhysA} and dense polymers \cite{denspol}. All of these
models display conformal symmetry at their critical points and the
proposed CFT corresponding to them is $c=-2$ model, which is a
logarithmic CFT\cite{LCFT}. Therefore, it is reasonable to seek a
straight way to connect ASM to this conformal field theory. One of
the first attempts was done by Mahieu and Ruelle. Through some
arguments of locality and scaling dimension and comparison of height
correlations of ASM with correlation functions of different fields
in  $c=-2$ model, they noted that certain fields can be associated
to a number of clusters; namely, WAC's \cite{MaRu}. Though it was a
great step forward, the method still had the shortcoming that it was
based only on comparison of correlation functions. Inspired by their
results and a method suggested by Ivashkevich to relate dense
polymers and CFT \cite{Ivashk} -- later elaborated for all trees and
forests by Caracciolo  {\it et al.}\cite{Sokal}-- Moghimi-Araghi,
Rajabpour and Rouhani solidly confirmed the correspondence between
$c=-2$ and ASM height correlations \cite{MRR}. They deduced the
results of Mahieu and Ruelle \cite{MaRu} in a straight-forward
manner and added some further subtleties to incorporate higher
corrections of height correlations in the fields.
\subsection{Grassmannian Method}
In their paper, Moghimi-Araghi, Rajabpour and Rouhani \cite{MRR}
re-expressed the Majumdar-Dhar probability of clusters in the bulk,
in terms of a contrived statistical system according to Berezin's
definition of Grassmannian integrals:
\begin{equation}
P(C)=\frac{\det\Delta^{'}_c}{\det\Delta}=\frac{\int d\t_i d\tb_j
\exp\left(\sum \t_i \Delta^{'}_{ij} \tb_j\right)}{\int d\t_i d\tb_j
\exp\left(\sum \t_i \Delta_{ij} \tb_j\right)}=\frac{\int d\t_i
d\tb_j \exp\left(\sum (\t_i \Delta_{ij} \tb_j+\t_i B_{ij}
\tb_j)\right)}{\int d\t_i d\tb_j \exp\left(\sum \t_i \Delta_{ij}
\tb_j\right)}.
\end{equation}
Here, $\Delta$ is the lattice Laplacian matrix and $B$ is the defect
matrix defined in ref. \cite{MajDharJPhysA}.

Evidently, since $[\theta_i\tb_j,\theta_k\tb_l]=0$ for all $i$, $j$,
$k$ and $l$, and hence, one can use Baker-Hausdorff-Campbell
formula, the above expression can be interpreted as the expectation
value of a field:
\begin{equation}\label{SCField1}
\varphi_c=\exp\left(\sum \t_i B_{ij} \tb_j\right).
\end{equation}

Subtleties arise in transition to the continuum limit; one can
perform the transition in different ways which yield different
results. Na\"{\i}vely, one can find the continuum limit of
$\Sigma\theta_i B_{ij}\tb_j$ in the exponent by expanding $\theta$
and $\tb$ around $\theta_{0}$. As an example, if we take the cluster
to be a single height-1 site, the corresponding field obtained with
this method turns out to be
\begin{eqnarray}\label{S0S1}
\sum \t_i B_{ij} \tb_j |_{S=S_0}  &\propto& \d \t \db \tb+
\db\t\d\tb -\frac{1}{4} \t\tb.
\end{eqnarray}
As $\theta$ and $\tb$ are Grassmann variables, the resulted
exponential can be calculated easily, since its Taylor series ends
at quadratic terms in $\t$ and $\tb$. The fields obtained in this
manner are not in accord with results of Mahieu and Ruelle
\cite{MaRu}, though there exist some similarities.

The second more careful method of transition to continuum is to
expand the exponential in terms of  $\theta$ and $\tb$ first, and
expanding  $\theta$ and $\tb$ around $\theta_{0}$ thereafter. As an
illustration, suppose we want to find out the field corresponding to
a height-1 site in the bulk \cite{MRR}. This can be sought through
the correlation function of two height-1 fields in the bulk in
Majumdar-Dhar method \cite{MajDharJPhysA} and relating that to the
fictitious statistical system:
\begin{eqnarray}\label{C1C2}
P(C_{1},C_{2})=\det(1+G\,B)&=&\frac{\int d\t_i d\tb_j
\exp\left(\sum (\t_i \Delta_{ij} \tb_j+\t_i B_{ij}^{1} \tb_j+\t_i
B_{ij}^{2} \tb_j)\right)}{\int d\t_i d\tb_j \exp\left(\sum \t_i
\Delta_{ij}
\tb_j\right)}\nonumber\\
\nonumber\\ &=&\left\langle\exp\left(\t_i B_{ij}^{1}
\tb_j\right)\exp\left(\t_i B_{ij}^{2} \tb_j\right)\right\rangle.
\end{eqnarray}
Here, $G$ is inverse of the matrix $\Delta$.

If we calculate $\langle\phi_{C_1}\phi_{C_2}\rangle$ using Wick
theorem, we find different terms. One can classify them according to
the number of long-range contractions between Grassmann variables of
the two fields - this number is always even, as a contraction of a
$\t$ of $C_1$ to a $\tb$ of $C_2$ must be compensated with a
contraction of a $\tb$ of $C_2$
with a $\theta$ of $C_1$. 
Having no long-rang contractions, simply reveals the probabilities
of the clusters individually and says nothing about correlation of
the two fields. The first relevant order is when we have two
long-range contractions and contract the other Grassmann variables
within their own clusters. So, to obtain the scaling field, on can
contract all the $\t$'s and $\tb$'s leaving a a pair of them
uncontracted. The resulting field is of the form $\phi=\sum \t_i
A_{ij}\tb_j$. Now, expanding  all $\theta$'s and $\tb$'s around
$\theta_{0}$ to the second order, one obtains the desired scaling
field; for instance, for one site cluster with height one, we have:
\begin{equation}\label{phis0}
\phi_{S_0}(z)= -\frac{4(\pi-2)}{\pi^{2}}:\!\d\t \db\tb + \db\t\d\tb
: .
\end{equation}
which is the same field obtained in ref. \cite{MaRu}. This expansion
is only up to the leading term in $a/r$, with $a$ and $r$ being
lattice spacing and and typical distance between two scaling fields.
In the next section, we consider higher order terms and derive
corrections to these scaling fields.

\section{Calculation of Higher Orders}
Obtaining the scaling fields in this way is a wearisome belabored
task and is prone to be afflicted by human error. Thus, we developed
a Mathematica \cite{Mathematica} code by which we could handle these
calculations fast, reliable and adjustable for different
configurations. We checked the code by recalculating the fields
assigned with simple WAC's; the final results were in accordance
with those of Moghimi-Araghi {\it et. al.} \cite{MRR}.

Now we would like to see what fields would arise if we consider
higher order terms in $a/r$. The configuration probability, $P(C)$
and the correlation function, $P(C_{1},C_{2})$, are the only
significant data we have at hand to figure out the form of the
corresponding scaling field. Following the Grassmannian method, one
can consider the expression $\exp\left(\t B\tb\right)$ as the field
associated with a WAC. First we expand this expression in terms of
$\t$ and $\tb$. In Grassmannian method we contract all the fields
except two of them, but we would like to derive several different
terms depending on how many $\t$'s and $\tb$'s are left
uncontracted. The resulting field will have the form:
\begin{equation}
\phi_C=c_0 {\bf I} + \sum \mathcal{A}_{ij}^{(1)}\t_i\tb_j +
\sum\mathcal{A}_{ijkl}^{(2)}\t_i\tb_j\t_k\tb_l+
\ldots+\sum\mathcal{A}_{i_1i_2\cdots
i_{2n}}^{(n)}\t_{i_1}\tb_{i_2}\ldots
\t_{i_{2n-1}}\tb_{i_{2n}}+\cdots,
\end{equation}
Note that the above series terminates due to Grassmann nature of
$\t$ and $\tb$. In each term $n$ is the number of uncontracted
pairs, and these pairs, when contracted with similar pair of another
field produce long-range correlations. To derive the coefficient
$\mathcal{A}^{(n)}$ one can use a generalized version of graphical
method used in \cite{MRR}; yet it is possible to find these
coefficients in an easier way. Contraction of all pairs will give us
the probability of the cluster which is $\langle\phi_C\rangle = P(C)
= \det\left(I+B\,G\right)$. It is easy to see that the coefficient
of $G_{ij}$ in $P(C)$ is $A_{ij}^{(1)}$ and the coefficient of
$G_{ij}G_{kl}$ is $\mathcal{A}_{ijkl}^{(2)}$. In this manner, we can
find out all the needed coefficients, $\mathcal{A}$'s, in $\phi_C$
easily; that is, firstly, we compute $P(C)=\det(I+G \,B)$ as a
function of $G_{ij}$'s. Secondly, we derive the coefficient of
$G_{ij}$, for all $i$ and $j$ to find $\mathcal{A}^{(1)}_{ij}$ or
derive the coefficient of $G_{ij}G_{kl}$ to find
$\mathcal{A}^{(2)}_{ijkl}$. Then remains just one additional step:
to expand $\t_i$'s and $\tb_i$'s around $\theta_0$. As an example,
doing all the procedure explained above we arrive at the following
expression for one-site height-one cluster up to $O(a^4)$:
\begin{eqnarray}\label{Final}
\phi_{S_0}=&a^2&
\frac{4(2-\pi)}{\pi^2}\left(\d\t\db\tb+\db\t\d\tb\right)\nonumber\\
+ &a^4& \frac{2(2-\pi)}{3\pi^2}\left(\d\t\d^3\tb+ \d^3\t\d\tb+
\db\t\db^3\tb+
\db^3\t\db\tb-\frac{3}{2(2-\pi)}\left(\d^2+\db^2\right)
\t\left(\d^2+\db^2\right)\tb\right)\nonumber\\
+&a^4& \frac{32(2-\pi)}{\pi} \,\d\t\d\tb\db\t\db\tb.
\end{eqnarray}

 In simplifying the original terms of
the field, the evolution equations of $c=-2$ action was applied to
eliminate some terms:
\begin{equation}\label{eq-mo}
\d\db\t=\d\db\tb=0.
\end{equation}
The first line of equation (\ref{Final}) is the ordinary field
derived before. The second line is the expansion of
$\mathcal{A}^{(2)}_{ij}$ up to $O(a^4)$, and the third line
originates from the term $\mathcal{A}^{(4)}_{ijkl}$. Note that there
are no terms of the order  $a^3$. In fact there were some terms, but
all vanish due to equation motion (equation (\ref{eq-mo})). The
fields of the order of $a^4$ are irrelevant under renormalization
group and so could be neglected in the scaling limit. But since in
ASM, size of lattice spacings may not be neglected in general, these
extra terms could be relevant in some calculations.

Though we have derived the higher corrections to the scaling field,
yet it is not the end of the story. If we are going to take care of
terms of the order of $a^4$ in scaling fields, we have to treat in
the same way with the action. The $c=-2$ action is obtained if we
collect only the terms up to second order of $a$. So, if the size of
the lattice is important, the action admits some modification.
Changing the action will change the Green function and we already
know that in calculation of higher order correction to correlation
functions, one has to take into account the higher order terms of
discrete Laplacian's Green function. This can be done readily in our
scheme: we have the discrete version of the action, just we have to
expand it up to order $a^4$. The result is:
\begin{equation}
A=\frac{1}{\pi}\int \left(\d\t\db\tb +a^2\left( \frac{1}{12}(
\d^2\t\d^2\tb+\db^2\t\db^2\tb)+\frac{1}{2}
\d\db\t\d\db\tb\right)\right).
\end{equation}
Note that the new action is still quadratic and hence integrable,
however it does not have conformal 
symmetry, although the off critical term is an irrelevant term and
vanishes under RG. But we are considering the problem when the smal
scale of the system can not be neglected completely, this forces us
to keep the extra, and irrelevant, terms in the action. The last
point we would like to mention is that the $a^3$ terms in the
scaling field still vanish, since the application of equation of
motion terms of order of $a^5$ only.

\section{Boundary Fields}

In this section we would like to derive the properties of scaling
fields in ASM when a boundary is present at the theory. This will
help us to derive finite size properties and surface critical
properties of the model. In the context of conformal field theory,
the boundary is usually taken to be the real line in the complex
plane and the system is supposed to fill the upper half-plane. The
correlation functions of scaling fields in this geometry reveals
surface critical exponents and the finite size scaling properties.
It is derived that under certain boundary conditions the
correlation functions of scaling fields in this geometry is the
holomorphic part of bulk correlation functions of same fields
together with their images (which live in the lower half
plane)\cite{Cardy84}. The idea was later generalized to the case
of logarithmic conformal field theory \cite{MR,KogWeat}.

Many properties of ASM in presence of such boundary has been
derived.  Ivashkevich calculated all two-point functions of all
height variables along closed and open boundaries
\cite{Ivashkevich94}. Using these correlation functions, Jeng
identified the height variables with certain fields in $c=-2$. A
more detailed treatment is given in \cite{JPR}. However, this
identification is through examining the correlation functions. We
would like to apply the Grassmaniann method to ASM with the
mentioned boundary. Note that as this method is only for WAC's, we
are not able to derive the height-2 or more one-site fields.

The question we are going to address is that if we consider the
model near a boundary, the toppling matrix, $\Delta$, and the Green
function $G$ change. Since in derivation of the fields using the
Grassmaninam method, we use Green function, the resulting field in
the presence of a boundary maybe different from the field in the
bulk.

To begin, we would like to calculate the one point function of a WAC
in the upper half boundary. This is given by the following
expression:
\begin{equation}
P(C)=\frac{\det\Delta^{'}_c}{\det\Delta}=\frac{\int d\t_i d\tb_j
\exp\left(\sum \t_i \Delta^{'}_{ij} \tb_j\right)}{\int d\t_i d\tb_j
\exp\left(\sum \t_i \Delta_{ij} \tb_j\right)}=\frac{\int d\t_i
d\tb_j \exp\left(\sum (\t_i \Delta_{ij} \tb_j+\t_i B_{ij}
\tb_j)\right)}{\int d\t_i d\tb_j \exp\left(\sum \t_i \Delta_{ij}
\tb_j\right)}.
\end{equation}
which is the same expression as in the bulk, just you should use the
appropriate matrix $\Delta$. Again $B^{C}$ is the matrix defined in
the Majumdar-Dhar method for the specified WAC, $C$. So it is seen
that the field associated with this cluster is again $\exp{\t_i
B^{C}_{ij} \tb_j}$, but we should keep in mind that to derive the
proper form of the field we use boundary Green functions which are
different from the bulk's Green functions. Let's expand this
expression and contract {\it all} the $\t$'s and $\tb$'s to find the
probability of the cluster. Note that as we would like to find
one-point correlation function, we do not leave two of the variables
uncontracted. The Green function of the theory with this geometry is
obtained easily by method of images:
\begin{eqnarray}
G_{\rm op}(\vec{r}_1,\vec{r}_2)=G(x,y_1-y_2)-G(x,y_1+y_2),\hspace{5mm}\nonumber\\
G_{\rm cl}(\vec{r}_1,\vec{r}_2)=G(x,y_1-y_2)+G(x,y_1+y_2-1),
\end{eqnarray}
where $G_{\rm op/cl}$ is the Green function open/closed boundary
conditions and $G$ is the Green function in the bulk. Also
$\vec{r}_i=(x_i,y_i)$ and $x=x_2-x_1$.

So, each contraction of $\t$ variables has two terms; one, which we
call it short range (SR) contraction, comes from the first terms in
above equations and is the bulk Green function of the points
$\vec{r}_1$ and $\vec{r}_2$. The other one, which we call it
image-long range (ILR) contraction, is the bulk Green function of
the point $\vec{r}_1$ with the image of the point $\vec{r}_2$ (see
figure 1).
\begin{figure}
\begin{picture}(200,170)(-80,120)
\includegraphics{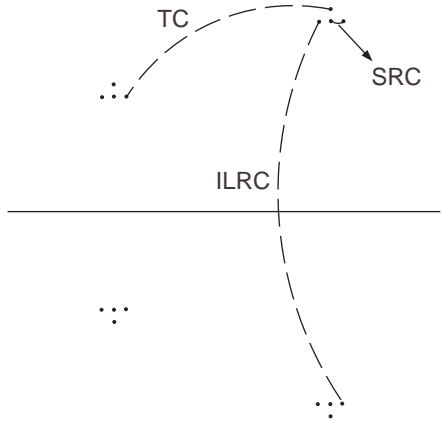}
\end{picture}
 \caption{Examples of SR-, ILR- and T-contractions near a boundary}
\end{figure}

Take a typical expression of $\t$'s and $\tb$'s to be fully
contracted. We can classify the terms appearing in the result by
number of ILR contractions (ILRC's), $n_I$. If $n_I$ is set to
zero, we will arrive at the bulk probability of the cluster $C$.
The first correction due to presence of the boundary appears when
$n_I=1$. This means that we should contract all the variables in
the way we did in the bulk except two which we will contract later
using a ILR contraction. The procedure reduces to the one
discussed by \cite{MRR} and for example the result for the
height-1 one-site cluster would be given by equation(\ref{phis0}),
only you have to contract the remaining $\t$ and $\tb$ using ILRC.
In the scaling limit this gives
\begin{equation}
\langle \phi_{S_0}(\vec{r})\rangle_{{\rm op/cl}}= P(1)\left( 1\mp
\frac{1}{4y^2}+\cdots\right),
\end{equation}
where $y$ is distance of the field from the boundary and minus/plus
sign corresponds to closed/open boundary condition which is
consistent with previous results. Also if we take a typical WAC,
with the corresponding field \cite{MaRu,Jeng05-1}
\begin{equation}
\phi_{S}=-\left[A_S :\d\t\db\tb+\db \t\d\tb:+
B_{1S}:\d\t\d\tb+\db\t\db\tb:+ iB_{2S}:\d\t\d\tb-\db\t\db\tb:
\right]
\end{equation}
it is easy to check that the second and third term of this field, do
not give any contribution when are ILR-contracted. Hence the
one-point boundary correlation function is obtained to be
\begin{equation}
\langle \phi_S \rangle_{{\rm op/cl}}= P(S)\mp \frac{A_S}{4y^2},
\end{equation}
just as it is indicated in \cite{Jeng05-1}.

Now we move on to calculate the field of a WAC explicitly. Suppose
you have two WAC's $C_1$ and $C_2$ and would like to compute the
probability of such configuration. Again we have an expression like
(\ref{C1C2}).  Now we can classify them depending on the number of
trans-contractions (TC's), $n_T$ and the number of ILRC's among the
inter-contractions, $n_I$. Setting both $n_T$ and $n_I$ equal to
zero, we will arrive at $P(C_1)P(C_2)$ with $P(C)$ being the bulk
probability of the cluster $C$. This term together with the term
coming from $n_T=0$ and $n_I=1$ reveals $\langle
\phi_{C_1}(\vec{r}_1)\rangle_{{\rm op/cl}}\langle
\phi_{C_2}(\vec{r}_2)\rangle_{{\rm op/cl}}$, the disjoint boundary
probability of the two clusters. To find the correlation of the two
clusters, we should set $n_T$ nonzero. Taking $n_T=2$ (the smallest
nonzero value for $n_T$) and $n_I=0$ the problem reduces to the one
in the bulk: contract all $\t$'s and $\tb$'s using {\it bulk} Green
function except two, which are left to be contracted with the $\t$
variables of the other cluster's field. This means that the field
derived in this manner is exactly the same as the one in the bulk
and we do not need to care about the fact that the Green functions
have changed.

The first terms that contain both joint probabilities and the effect
of boundary Green function, appears when we set $n_T=2$ and $n_I=1$.
This means that you should contract all $\t$'s $\tb$'s but four,
using SRC, two of remaining will be contracted with the variables of
other fields and two will be ILR contracted. Such field has been
calculated in the previous section and we have seen it is of the
order of $a^4$ and vanishes in the scaling limit. So, if we are
considering the problem in the scaling limit, the Grassmaniann
method says the the boundary field is just the same as bulk field.
However, this does not mean that the effect of boundary could be
neglected completely, because in the trans-contractions should use
boundary Green functions. \vspace{5mm}

\noindent {\Large \bf Acknowledgment}

We would like to thank S. Rouhani and M. A. Rajabpour for their
helpful comments and careful reading of the manuscript. We thank an
unknown referee pointed the need to discuss the extra term appearing
in the action.

\end{document}